# 3D Printing Neutron Detectors using Scintillating BN/ZnS Resin

**P. Stowell*[a], Z. Kutz[b], S. Fargher[a], and L. F. Thompson[a]**

[a] *University of Sheffield,
   Sheffield, United Kingdom*
[b] *Technical University Berlin,
   Berlin, Germany*
   *E-mail*: p.stowell@sheffield.ac.uk

ABSTRACT: In this paper we demonstrate that it is possible to produce low cost neutron-sensitive detectors using stereo-lithography additive manufacturing. A curable scintillating resin is made by mixing BN/ZnS with a commercially available UV resin. This resin is used to print several small area neutron detectors made of arrays of BN/ZnS cones that can be directly coupled to a photo-multiplier tube.

KEYWORDS: 3D Printing; Neutron Detection; Radiation Detection.

# Contents

## 1. Introduction

Helium 3 detectors are considered the "gold standard" for neutron detection due to their high efficiency and insensitivity to gamma radiation. Driven by shortages of Helium-3 in recent years however, there has been rapid development in alternatives to this technology using solid state detectors or scintillating materials. Li/ZnS composite materials are one such alternative that are sensitive to thermal neutrons. A neutron capture by Li results in a long characteristic decay time of the ZnS scintillator pulse which can be easily identified using pulse shape discrimination techniques. Since Li/ZnS is opaque, the sensitivity of a detector is governed by the total active surface area coupled to a photo-detector. Several designs have made use of rippled structures, or light guides to effectively increase the total sensitive area of these detectors with a minimal number of photo-detectors [1][2]. In [3], it was shown that unenriched hexagonal BN mixed with ZnS was a suitable lower cost alternative to Li/ZnS provided large area, low cost detectors could be constructed using wavelength shifting light guides for readout.

Recently, with rapid advances in the availability and reduction in cost of 3D printing, there have been several studies on the development of 3D printing plastic scintillator [4][5][6]. In [5], scintillating dopants are mixed with a photopolymer resin to produce a solution that can be cured (solidified) using 385 nm light. This makes it possible to use StereoLithography (SL) to produce 3D objects by curing multiple 2D images from the resin and stacking them on top of one another as shown in Figure 1. A high efficiency scintillator resin that can be 3D printed using SL has the potential to radically transform the radiation detector field, making it possible to prototype complex structures that are not feasible using standard manufacturing techniques.

In this paper we demonstrate that it is possible to 3D print complex, neutron-sensitive geometries by mixing unenriched BN/ZnS with a UV curing resin. Since only the top surface layers of a BN/ZnS or Li/ZnS detector are sensitive to neutrons due to the material's opaqueness, it is feasible to mix these compounds into commercially available UV curing resins to produce sensitive detector prototypes at low cost. This paper is organised as follows; in section 2 the scintillating resin mixture and curing procedures are discussed, in section 3 the scintillator testing method is described, in section 4 the pulse discrimination performance for three differing geometries is evaluated, before finally section 5 discusses the conclusions and potential uses for this technique.

## 2. UV Cured BN/ZnS Scintillator

To produce a neutron sensitive scintillating resin, BN and ZnS powders were mixed into a commercially available clear UV resin from AnyCubic Photon Ltd [7] with a weight ratio of

$$1 \text{ BN} : 5 \text{ ZnS} : 5 \text{ Clear Resin}.$$

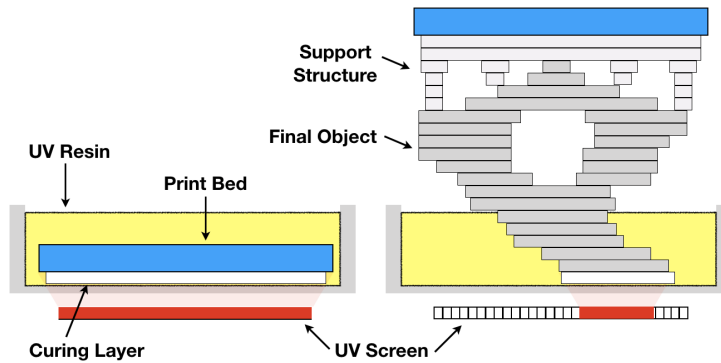

Figure 1. Stereolithography 3D printing method. Objects are "sliced" into multiple 2D layers which are cured by a UV screen projected into the resin container. 3D objects are constructed by curing many different 2D layers on top of one another, moving the print bed up slightly after each layer.

In [3], ratios greater than 1:3 of BN to ZnS were shown to satisfy the condition that each BN molecule is sufficiently surrounded by ZnS that the neutron capture products lead to a strong scintillation signal. A slightly larger conservative fraction of ZnS was chosen in this work to ensure this was still the case in the UV curable resin mixture. To ensure adequate mixing of the BN/ZnS compound, these were first mixed dry inside a blacked out container.

The resulting powder was then placed on a slow magnetic stirrer as UV resin was poured in to the blackout container to produce an opaque white resin. This scintillating resin was slowly stirred for two hours at room temperature to ensure a good solution with no air bubbles. The resin was kept in the original blackout container throughout the entire mixing process. This minimised unwanted curing of the resin before the printing stage, as this was found to prematurely thicken the mixture making it difficult to print with. Ideally, the ratio of BN/ZnS to resin should be as high as possible as the resin is insensitive to neutron interactions, however initial tests found that mixtures with a ratio much higher than 1:1 began to behave as a non-Newtonian fluid that made it difficult to produce a well-mixed solution of BN/ZnS/resin that could be printed with.

An AnyCubic Photon SL printer was used to construct 3D objects from this scintillating resin, curing with a 405 nm UV screen [8] with voxel resolution of ~10um. When trying to print with the scintillating resin using the standard curing settings suggested for the original clear resin, many prints failed as the resin tended to cure directly to the UV screen instead of the print bed. This was determined to be due to poor UV light transport through the opaque resin; since insufficient light could pass through, the first few layers were only slightly cured far away from the UV screen. This was solved by printing the first few layers of disposable support structures using a standard clear UV resin before filling the resin container with BN/ZnS/resin mixture once a stable base on the print bed had been formed.

This dual resin method made it possible to print 3D objects, however as shown in Figure 2, it was still not possible to produce high quality outputs due to the print bed developing a skew over the course of a print. This was found to only occur when using vertical layer sizes < 50 μm. It is believed this is a result of the slightly non-newtonian behaviour of the resin when mixed at a 6:5 powder to resin ratio. This behaviour leads to a minimum compression thickness below which excess pressure is applied to the print bed that leads to a small skew after each layer is cured. This results in failed print parts with noticeable warping.

Changing to a minimum vertical layer size of 50 μm and avoiding printing objects with large surface areas on consecutive layers reduced the forces on the print bed and made it possible to successfully produce complex prints from the resin. Figure 2 shows a successfully printed neutron detector made of an array of 3 mm radius BN/ZnS cones that can be directly mounted onto the face of Photo-Multiplier Tube (PMT) for readout. This cone structure makes it possible to couple a larger surface area to a PMT compared to a flat BN/ZnS sheet simply by changing the geometry of the printed part. Since the minimum compression thickness is believed to be a function of the powder to resin ratio, it is possible that reduced mixing ratios could be used to produce parts with a much finer vertical resolution in the future albeit with lower detection efficiencies.

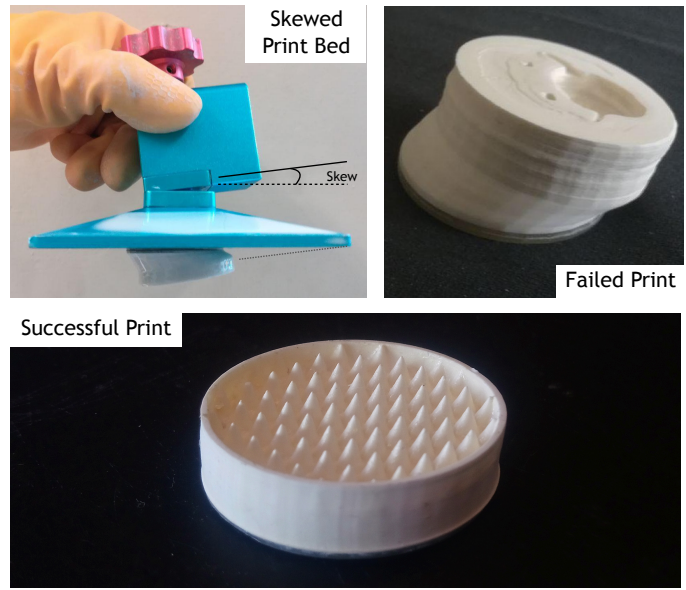

Figure 2. (top left) Skewed print bed due to the BN/ZnS UV curing resin due to using too small vertical layer heights when printing. (top-right) Example of a failed, skewed print. Modified curing settings with thicker printer layers lead to successful prints, allowing more complex geometries to be produced (bottom).

## 3. Scintillator Testing

To evaluate the neutron discrimination power of this 3D printed cone detector, a test stand was setup inside a dark box as shown in Figure 3. The neutron detector was mounted directly onto the face of an ET-Enterprises 9902B PMT. The PMT was supplied with a 1.2kV High Voltage (HV) bias using an ET-HV3820AN base. The output pulse from the PMT was fed directly into a DRS4 Evaluation Board, which digitised the signal into 14 ADC-bits at a rate of 1GHz with a 1024 sample acquisition window. Since this digitiser does not have a Pulse Shape Discrimination (PSD) trigger built in, a fixed threshold discriminator set at 50mV was used to trigger the digitisation of a full pulse which was saved for further processing offline.

In early tests it was noted that the ZnS was easily activated by UV light, therefore each 3D print was kept in a dark box for 24 hours before being tested. In each test, the BN/ZnS detector was removed from storage and mounted inside the test stand before being left for 10 minutes for any stray activation of the ZnS during its installation to decay. Following this installation period, a one hour data taking period was started on the digitiser. To provide an source of neutrons, a Cf-252 source (0.729MBq) was placed outside the dark box surrounded by a HDPE moderator, approximately 40cm away from the detector. As expected, neutrons from the Cf-252 source were found to produce long pulses in the BN/ZnS mixture, of the order of several microseconds. In contrast pulses obtained either with no source present, or with a Cs-137 gamma source (213MBq), were found to have widths below 30ns. Example pulses from a characteristic neutron and background pulses can be seen in Figure 4.

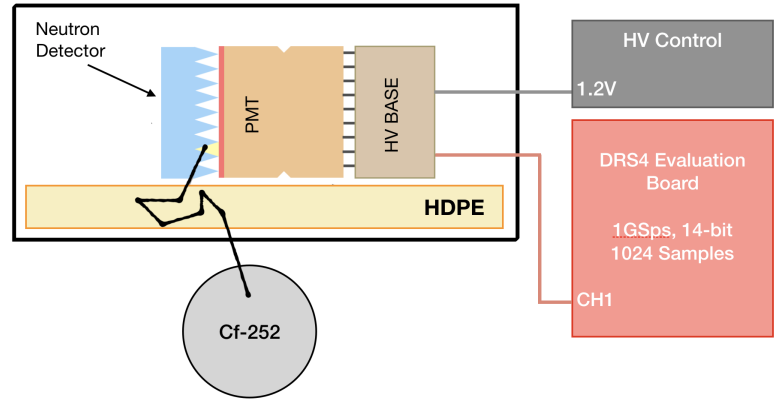

Figure 3. Photo-multiplier test stand layout. A neutron capture on the BN/ZnS cone detector results in a long scintillator pulse in the PMT which is tagged during offline analysis of the DRS4 digitiser data.

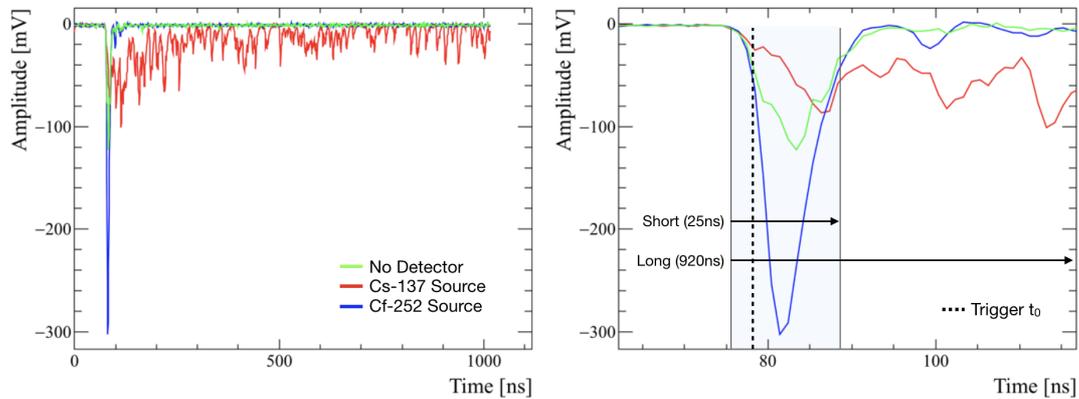

Figure 4. Characteristic difference in pulse decay times for neutrons, gamma and PMT-noise events for the detector. The long decay time of neutron events are used to discriminate neutrons from background signals by comparing short and long integrals.

Using the digitised pulses collected over each one hour period, a Pulse Shape Discrimination (PSD) algorithm was applied to the pulses offline. A PSD Ratio metric was defined as the ratio of "short' and "long" integrals. Given a trigger time, $T_0$, the "short" integral was defined as the total pulse area from $T_0$-6 ns to $T_0$+20 ns, and "long" was the total area from $T_0$-6 ns to $T_0$+920 ns. A width of 26 ns for the short integral was chosen based on studies of test pulses taken with either Cf-252 or Cs-137 sources present, and data obtained when no neutron detector was attached to the PMT face. To obtain a high neutron detector efficiency, it was necessary to operate the PMT with high gain, as typically only several photons were detector every few nano-seconds during a neutron capture event. The drawback of this is that the PMT noise typically has a high amplitude and is too difficult to separate from gamma interaction events based on its pulse height or timing information. Despite this, the long characteristic decay time of neutron capture events in the BN/ZnS made it possible to isolate these events from the PMT background and gamma events.

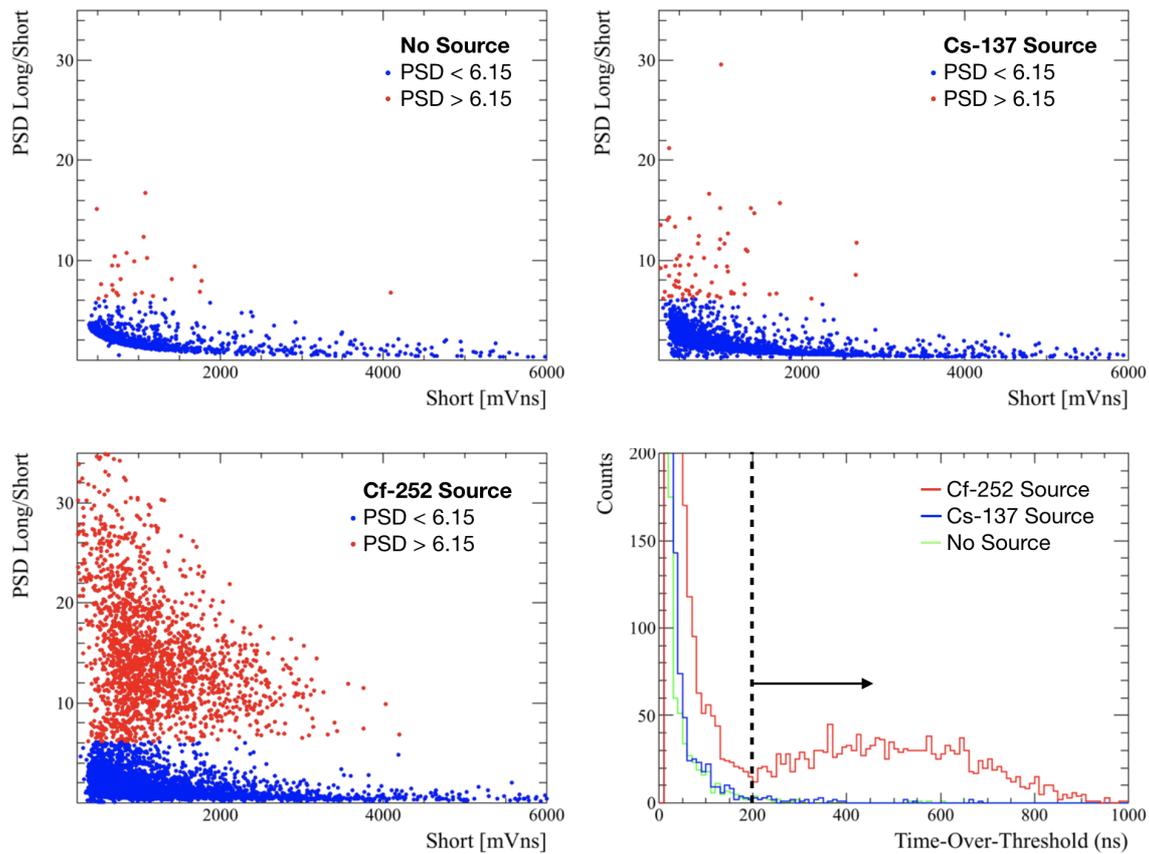

Figure 5. Pulse shape discrimination results for a UV-cured neutron detector. (top-left) No source present. (top-right) Cs-137 gamma source present. (bottom-left) Cf-252 present. (bottom-right) Alternative time-over-threshold discrimination metric performance for various source conditions. All Cf-252 distributions show a clear excess of events due to neutrons capturing in the detector.

Plotting these PSD distributions as a function of the short pulse integral in Figure 5 shows a distribution of background events at PSD Ratio < 6.15 that is present even when no source is nearby. When a Cs-137 gamma radiation source is added, a significant increase in the number of these events is observed, but it is difficult to separate the two types of events based on pulse height or timing information. In contrast, when a Cf-252 source is placed outside the dark box, a clear excess of events with long decaying pulses appears above PSD Ratio > 6.15. In addition to this clear excess, additional events are also observed in the PSD Ratio < 6.15 region when a Cf-252 source is present. Due to Cf-252 gamma radiation events, and additional false triggers that are generated from events where an additional event occurred detected during the readout dead time of the DRS4 evaluation board (1ms), and the trigger misinterprets the part of the long scintillation decay as a new trigger event. It is expected that a customised data acquisition system could be used to further optimise the efficiency of these detectors, however the DRS4 board has been shown to achieve a satisfactory level of neutron detection efficiency, given the relatively low cost to produce the neutron sensitive detectors themselves. Furthermore, given the long decay time of the detectors when exposed to a neutron source, it was possible to use a simplified discriminator metric such as Time-Over-Threshold as shown in Figure 5 to reliably identify neutron events, meaning they could be used with a single channel comparator to produce an extremely low cost neutron detector.

## 4. PSD Performance Evaluation

This novel development in 3D printed structures makes it possible to investigate alternative ways to print optimised structures. In this section we briefly investigate the possible effect 3D printed cone size could have on detector performance for a neutron detector similar to that shown in the previous section. By printing narrower conical structures with the same height, the total sensitive area of a neutron detector that emits light from its surface isotropically should be effectively increased. To investigate the effect this could have on discrimination power, two alternative conical geometries were investigated. Conical radii of 2 mm (small), 3 mm (medium), and 4 mm (large) were considered as shown in Figure 6. The approximate increase in surface area compared to a flat disk when neglecting the sensitive surface of one cone being blocked by another, is 401% (small), 273% (medium), and 211% (large) respectively. Unfortunately, due to the bed skew issues discussed previously, it was not possible to print a flat neutron sensitive disk for reliable comparison, however relative comparisons of such geometries can still highlight whether specific optimisation methods can improve neutron detection or light collection efficiency.

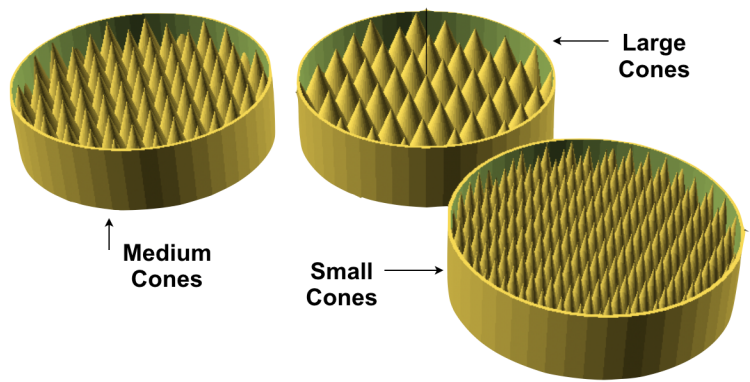

Figure 6. Different cone detector geometries considered in this study.

For each geometry study, the analysis procedure described in Section 3 was repeated. Digitiser pulses were recorded for a one hour period, with and without a Cf-252 source inserted into the PMT test stand, before being analysed offline. As can be seen in Figure 7, the addition of the Cf-252 source produces a clear excess of events around PSD ~ 12 for short pulse integrals greater than 800 mVs due to neutron interactions in the BN/ZnS. No significant difference in the shape or position of this distribution was observed when changing the detectors cone size as shown in Figure 7.

Comparisons of the total number of events detected with PSD Ratio > 6.15 however suggest a very slight change in total detection efficiency, with the small cones showing a $+6.6 \pm 3.2\%$ increase in efficiency, and the large cones having a $-8.2 \pm 3.4\%$ reduction in efficiency when compared to the medium cones detector. Large increases in the number of events observed were also observed for the small cones detector even with no source present, suggesting that some fraction of the variation in events observed for PSD < 6.15 was due to the surface of the BN/ZnS activating when exposed to light when the detectors were installed in the dark box. One possible reason for such small differences in total detected neutron events compared to that expected from the change in total surface area is due to the voxelised nature of the 3D printed cone arrays. When considering perfectly smooth cones there is a large effective increase in area with reducing cone diameter. In a voxelised geometry however, all surfaces are either directly parallel or perpendicular to the PMT face when imaged at a 10 μm scale. Therefore the increase in effective area is likely to be significantly reduced. Despite this lack of improvement in discrimination power, these studies have demonstrated that the UV curing resin suggested in this paper can be used to produce extremely fine scale features without further degradation of the detector sensitivity.

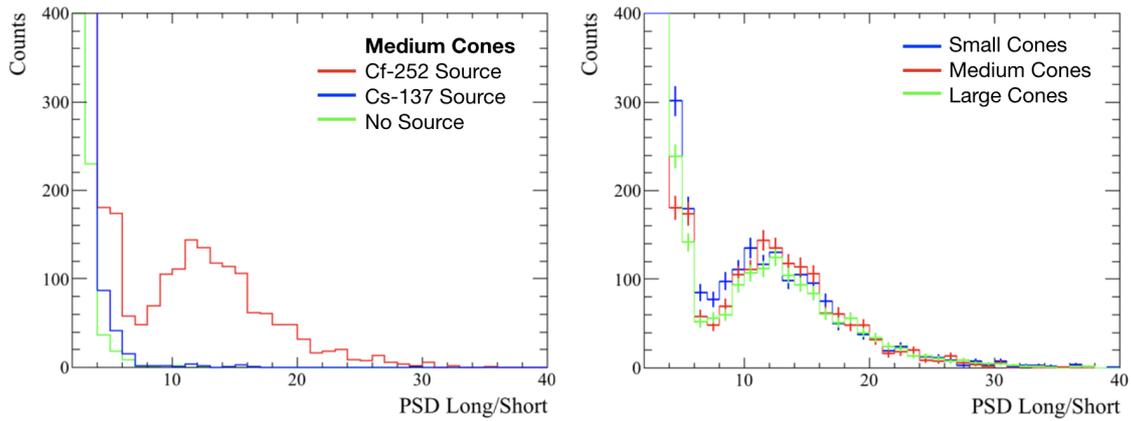

Figure 7. PSD Ratio comparisons for neutron detectors with variable cone dimensions for short pulse integrals > 800 mVs. All pieces were found to be sensitive to a Cf-252 source, but showed no significant variation in pulse height or discrimination power. (left) Comparison of the results obtained with the Medium Cones detector under different radiation field conditions to highlight the neutron events excess. (right) Comparison of the neutron event excess for the three different candidate detector geometries.

Table 1. Detected events in different PSD regions for the three neutron detector geometries investigated in this study.

| Detector | No Source | | Cf-252 Source | | Cs-137 Source | |
|---|---|---|---|---|---|---|
| | PSD < 6.15 | PSD > 6.15 | PSD < 6.15 | PSD > 6.15 | PSD < 6.15 | PSD > 6.15 |
| Small | 2840 ± 53 | 23 ± 5 | 13454 ± 115 | 2152 ± 46 | - | - |
| Medium | 1421 ± 38 | 25 ± 5 | 9345 ± 97 | 2019 ± 45 | 10319 ± 101 | 51 ± 7 |
| Large | 1351 ± 37 | 16 ± 4 | 7482 ± 86 | 1852 ± 43 | - | - |

5. Conclusions

This work has demonstrated that it is feasible to produce novel neutron detectors using additive manufacturing by mixing BN/ZnS compounds into commercially available UV resin. This makes it possible to produce complex structures that could be used to optimise detector sensitivity. Whilst changes in total surface area in a printed cone array geometry were found to not significantly improve the pulse shape discrimination power, alternative geometries produced on printers with finer resolution could improve this in the future.

The use of low cost clear resin to print support structures was found to not only aid in print stability but also reduce the total cost of the prints. Custom built 3D printers could extend this further by using in-situ resin mixing to produce complex structures with varying levels of scintillator doping across a part. This could potentially provide positional information, or improved sensitivity and discrimination capability in mixed radiation fields.

## Acknowledgments

This work was supported by a DAAD Rise Worldwide summer student grant. We would also like to thank John McMillan in supporting this work by providing BN/ZnS raw materials to construct these detectors.